\documentclass{article}

\usepackage[preprint]{neurips_2024}

\usepackage[utf8]{inputenc} % allow utf-8 input
\usepackage[T1]{fontenc}    % use 8-bit T1 fonts
\usepackage{hyperref}       % hyperlinks
\usepackage{url}            % simple URL typesetting
\usepackage{booktabs}       % professional-quality tables
\usepackage{amsfonts}       % blackboard math symbols
\usepackage{nicefrac}       % compact symbols for 1/2, etc.
\usepackage{microtype}      % microtypography
\usepackage{xcolor}         % colors
\usepackage{amsmath}
\usepackage{graphicx}       % graphicx for figures
\usepackage{enumitem}

\title{Investigating Group Relative Policy Optimization for Diffusion Transformer based Text-to-Audio Generation}

\author{%
    \textbf{Yi Gu}$^1$\thanks{Work done during an internship at Microsoft speech team.} \ \ 
    \textbf{Yanqing Liu}$^2$\thanks{Corresponding author: yanqliu@microsoft.com} \ \ 
    \textbf{Chen Yang}$^1$ \ \ 
    \textbf{Sheng Zhao}$^2$ \\
    $^1$Tsinghua University\\
    $^2$Microsoft
}

\begin{document}

\maketitle

\begin{abstract}

Text-to-audio (T2A) generation has advanced considerably in recent years, yet existing methods continue to face challenges in accurately rendering complex text prompts—particularly those involving intricate audio effects—and achieving precise text-audio alignment. While prior approaches have explored data augmentation, explicit timing conditioning, and reinforcement learning, overall synthesis quality remains constrained. In this work, we experiment with reinforcement learning to further enhance T2A generation quality, building on diffusion transformer (DiT)-based architectures. Our method first employs a large language model (LLM) to generate high-fidelity, richly detailed audio captions, substantially improving text-audio semantic alignment, especially for ambiguous or underspecified prompts. We then apply Group Relative Policy Optimization (GRPO)—a recently introduced reinforcement learning algorithm—to fine-tune the T2A model. Through systematic experimentation with diverse reward functions (including CLAP, KL, FAD, and their combinations), we identify the key drivers of effective RL in audio synthesis and analyze how reward design impacts final audio quality. Experimental results demonstrate that GRPO-based fine-tuning yield substantial gains in synthesis fidelity and prompt adherence.

\end{abstract}

\section{Introduction}

Text-to-audio (TTA) generation holds significant promise for content creation and artistic applications, from powering podcasts and storytelling platforms to enabling speech-driven video synthesis and immersive sound design. However, generating coherent, prompt-aligned audio—particularly for complex or lengthy prompts—has historically required intensive data labeling and precise text-audio alignment. Recent advancements in TTA models \citep{zheng2025picoaudio2,jiang2025controlaudio,hung2024tangoflux,jiang2025freeaudio,evans2025stable,guan2024lafma,liu2024audiolcm,mo2024text} have introduced innovative techniques, such as latent consistency models for rapid inference with minimal sampling steps \citep{liu2024audiolcm}, flow-matching paradigms for accelerated and high-fidelity generation \citep{guan2024lafma,hung2024tangoflux}, training-free timing planning for long-form controllability \citep{jiang2025freeaudio}, progressive diffusion for decoupled temporal and semantic control \citep{jiang2025controlaudio}, timestamp- or natural-language-based temporal guidance \citep{zheng2025picoaudio2}, and open-source architectures optimized for variable-length sound effects and music \citep{evans2025stable}. These developments have markedly improved diversity, expressiveness, inference efficiency, long-form synthesis, and fine-grained controllability, facilitating applications like synchronized audiovisual content and extended narrative audio. Despite these advances, a significant gap remains: achieving high-fidelity audio with robust semantic and temporal alignment is still challenging for intricate or extended prompts involving multiple sequential events, subtle nuances, or prolonged narratives. Existing methods frequently struggle to capture fine-grained textual details, resulting in event omissions, hallucinations, misordered sequences, or weakened text-audio correspondence—particularly in real-world, long-form scenarios.

We attribute the aforementioned limitations primarily to deficiencies in data quality and modeling approaches. Data quality plays a critical role in the controllability and fidelity of text-to-audio generation systems, where poor-quality captions frequently lead to incorrect audio events, disordered temporal sequencing, or semantic misinterpretations of the input prompt. To address these issues, we propose enhancing prompt-following capability by augmenting the training data with higher-quality, more detailed textual descriptions. Specifically, we first design a set of carefully engineered prompts that elicit rich, accurate, and structurally consistent audio captions from large language models (LLMs). These prompts are then applied to our base audio dataset, leveraging powerful LLMs such as GPT-4 \citep{openai2023gpt4}, DeepSeek-V2 \citep{deepseek2024v2}, and Qwen-2.5 \citep{qwen2025qwen25} to generate enhanced descriptions automatically. Finally, we fine-tune text-to-audio models on this augmented dataset, resulting in substantially improved alignment, controllability, and overall synthesis quality. We then apply reinforcement learning, specifically Group Relative Policy Optimization (GRPO), with a diverse set of reward functions (including CLAP, KL, FAD, and their weighted combinations) to further enhance text-to-audio (T2A) generation quality. Through systematic experimentation, we analyze how different reward designs influence key aspects of T2A synthesis, including distributional fidelity, sample-level alignment, semantic prompt adherence, and overall audio quality.

Pretraining of text-to-audio (T2A) models—whether autoregressive or diffusion-based—commonly results in imprecise text-audio alignment. In contrast, post-training reinforcement learning techniques have proven highly effective in large language models (LLMs) for mitigating alignment mismatches. The substantial progress in LLMs owes much to alignment stages, typically using Reinforcement Learning from Human Feedback (RLHF) or similar reward-based optimization methods, which align model outputs closely with human preferences. However, this alignment paradigm remains underexplored in T2A development, largely due to a key tooling gap: the scarcity of reliable, large-scale preference pair datasets. Unlike the LLM domain, which benefits from established reward models and abundant human feedback resources, the T2A field currently lacks standardized, scalable approaches for collecting and utilizing such preference data, hindering effective post-training alignment. In this work, we address this limitation by exploring reinforcement learning-based alignment for T2A generation. Specifically, we investigate dataset preparation strategies for RL training and systematically examine how different reward function designs (including CLAP, KL, FAD, and their combinations) influence final audio quality. Our goal is to reduce alignment mismatches beyond what data augmentation alone can achieve, providing insights into effective reward modeling for text-to-audio synthesis.

In this work, we investigate reinforcement learning for text-to-audio (T2A) generation, conducting systematic experiments to uncover promising directions for future RL-based T2A modeling. Our approach encompasses three key aspects: high-quality data construction, model training via reinforcement learning, and reward function design. In the data construction stage, we start with a large-scale collection of ⟨text, audio⟩ pairs and augment it by employing large language models (LLMs) to produce richer, more detailed, and semantically precise captions. For the reinforcement learning stage, we adopt Group Relative Policy Optimization (GRPO) and systematically explore a range of reward functions—including CLAP score, Fréchet Audio Distance (FAD), Kullback–Leibler (KL) divergence, and their weighted combinations—demonstrating consistent improvements in generation quality across different reward formulations. These findings offer valuable insights into effective reward engineering for enhancing text-audio alignment and overall synthesis fidelity in T2A systems.

In summary, we make the following contributions. 
\begin{itemize} 
\item We augment the text descriptions using a large language model (LLM) to produce richer and more detailed captions. This enhancement improves the robustness of rendered audio effects—particularly for complex or subtle acoustic phenomena—and we compare these gains with the improvements achieved through reinforcement learning.
 \item We further enhance text-to-audio (T2A) generation quality by applying Group Relative Policy Optimization (GRPO) to a diffusion transformer (DiT)-based T2A model. We systematically analyze performance differences across various reward functions and conduct extensive ablation experiments to verify the contribution of each component to improvements in benchmark scores.
\end{itemize}

\section{Related work}

\subsection{Generative Speech Models}

Generative speech models based on autoregressive or diffusion architectures have significantly advanced end-to-end speech quality across tasks such as text-to-speech (TTS), speech recognition, speech-to-speech conversion, and multimodal modeling \citep{hu2026qwen3,xu2025qwen3, wu2025step,wang2023neural,chen2024vall,zhang2023speak,shen2023naturalspeech,kanda2024making, liu2025next,ju2024naturalspeech,han2024vall,eskimez2024e2,li2024investigating,yang2025pseudo,wang2025felle,li2018close,li2019neural,li2020robutrans,liu2021delightfultts,tan2024naturalspeech,yuan2024continuous,eskimez2024e2,meng2024autoregressive,liu2022delightfultts}.Continuous speech features—such as mel-spectrograms derived from signal processing method or discrete speech tokens obtained via residual vector quantization (RVQ)-based codec models—have frequently served as effective intermediate representations in speech modeling, demonstrating strong performance in TTS and speech-to-speech tasks. These successes in speech synthesis have inspired similar architectural, data-scaling, and training strategies for text-to-audio (T2A) generation, which has emerged as a prominent recent research trend.

\subsection{Text-to-Audio}

Text-to-audio (T2A) generation has emerged as a prominent field, enabling models to reliably synthesize audio that is semantically consistent with given text descriptions. The domain now predominantly relies on diffusion and flow-matching models, which iteratively denoise pure noise in the latent space conditioned on text embeddings and time steps. Diffusion-based approaches \citep{yang2023diffsound,liu2024audioldm,evans2025stable,majumder2024tango,novack2025fast,liu2025flashaudio,feng2025audiospa,jeong2025read,ronchini2024synthetic,lee2024challenge,niu2024soundlocd,jiang2025freeaudio, zhao2025audioturbo,wang2025t2a,yuan2025dreamaudio,vora2025ptq4adm,tian2025audiox} have become the dominant paradigm, owing to their superior audio fidelity and generation efficiency. Recent research has increasingly focused on long-form generation and precise time-aligned audio synthesis. Stable Audio \citep{evans2025stable} advanced long-form capabilities by training diffusion models on large-scale proprietary datasets and incorporating timing-conditioned embeddings, though these primarily regulate overall duration rather than providing fine-grained control over specific event timing. TANGO and TangoFlux \citep{majumder2024tango,hung2024tangoflux} improved compositional accuracy and alignment with complex prompts through instruction-tuned language models and preference-based fine-tuning. Despite these advances, achieving precise, event-level text-audio alignment within diffusion frameworks remains a critical challenge. Several recent efforts have directly addressed this control gap. ControlAudio \citep{jiang2025controlaudio} pretrains a Diffusion Transformer (DiT) on large-scale text-audio pairs and integrates timing and phoneme features via unified semantic representations, enabling scalable and fine-grained controllability. PicoAudio2 \citep{zheng2025picoaudio2} advances data-centric methods by employing a grounding model to annotate event timestamps in real audio-text datasets, thereby curating a strongly temporally aligned corpus that augments existing synthetic data. Finally, FreeAudio \citep{jiang2025freeaudio} introduces a training-free T2A approach that simultaneously achieves precise timing control and coherent long-form synthesis from complex natural language prompts, marking a significant step toward structurally consistent and controllable audio generation.

\subsection{Group Relative Policy Optimization (GRPO)}

Reinforcement learning (RL) has emerged as a key paradigm for aligning large language models (LLMs) across modalities, including text, image, and—most recently—speech. The lightweight, policy-based method Group Relative Policy Optimization (GRPO) \citep{shao2024deepseekmath} has proven particularly effective in LLM post-training, enhancing both safety and utility. Building on this success, recent studies have explored RL's potential in speech processing tasks, including Automatic Speech Recognition (ASR) and Text-to-Speech (TTS). For ASR, \citep{gao2025explore} applied the GRPO framework, experimenting with various rule-based reward functions and examining the impact of RL data construction strategies. In TTS, \citep{gao2025differentiable} introduced Differentiable Reward Optimization (DiffRO), which directly computes rewards based on neural codec tokens, significantly improving pronunciation accuracy in TTS systems. Extending ASR research, \citep{liu2025group} proposed a more informative composite reward function that integrates Character Error Rate (CER) with Negative Log-Likelihood (NLL) from the ASR model, delivering more accurate and stable reward signals.

\section{Method}
\label{gen_inst}

We adopt the pretrained TangoFlux model \cite{hung2024tangoflux} as our experimental baseline. TangoFlux is built upon Diffusion Transformer (DiT) and Multimodal Diffusion Transformer (MMDiT) architectures and generates high-fidelity 44.1 kHz audio clips up to 30 seconds in duration. It learns a rectified flow trajectory in the continuous latent space of audio—encoded via the variational autoencoder (VAE) from Stable Audio Open—while conditioning on text representations from FLAN-T5. To construct reinforcement learning experiment, we first construct an enriched dataset with more detailed and semantically precise text descriptions, which then serves as the training set for subsequent reinforcement learning.

\subsection{Descriptions Augmentation}

Obtaining diverse and high-quality training data remains a major challenge in developing high-performance text-to-audio (T2A) generation models. Existing labeled audio datasets are typically limited by short, low-diversity, and occasionally inaccurate captions. While human annotators can produce detailed and reliable descriptions, manual annotation is expensive and difficult to scale. To construct an effective training set for reinforcement learning experiments, we employ a data augmentation pipeline that leverages large language models (LLMs) to generate rich, detailed audio descriptions from sparse original tags and short captions. Our approach consists of the following steps: We design multiple prompting rule sets engineered to elicit longer, more descriptive, and semantically faithful outputs from the LLM. We systematically evaluate these rule sets across several LLMs using a small, human-verified evaluation subset. Human annotators score the generated descriptions for quality and fidelity, and we select the highest-performing configuration based on average scores. Once the optimal prompt template and LLM are determined, we apply them at scale to the full collection of original tags and labels, producing a substantially enriched corpus of detailed textual descriptions. During training, we combine the newly generated descriptions with the original captions by randomly sampling from both sources according to a controllable mixing ratio. Model performance is continuously monitored on a held-out validation set to identify the most effective ratio and to avoid overfitting to synthetic descriptions. This pipeline enables scalable, high-quality data augmentation that significantly enhances prompt adherence and overall synthesis fidelity in subsequent reinforcement learning stages.

\subsection{ Flow-Matching with Group Relative Policy Optimization}

%\paragraph{Flow-Matching}

\subsubsection{ Flow-Matching }

Diffusion-based generative models are highly sensitive to the choice of noise scheduler; suboptimal schedules can substantially degrade performance. In contrast, the flow matching (FM) framework \citep{lipman2022flow} demonstrates exceptional robustness to hyperparameter variations and has recently achieved state-of-the-art results across diverse audio generation tasks, including text-to-audio (T2A), text-to-speech (TTS), and speech-to-speech conversion. Flow matching learns a time-dependent velocity field that directly transports samples from a simple base distribution (typically Gaussian noise) to the target data distribution. This enables straight-through training via a straightforward velocity regression objective and supports exact likelihood computation through ordinary differential equations (ODEs). The paradigm has also driven recent breakthroughs in text-to-image generation, yielding photorealistic images of unprecedented fidelity. Furthermore, reinforcement learning (RL)—which previously delivered transformative gains in autoregressive language modeling—has recently been successfully integrated with flow matching objectives, yielding substantial additional improvements in controllable and expressive text-to-speech synthesis \citep{sun2025f5r,liu2025flow}.

\subsubsection{Group Relative Policy Optimization}

Reinforcement learning (RL) has driven substantial progress in aligning large language models (LLMs) with human preferences. Preference-based optimization methods such as Direct Preference Optimization (DPO) and Group Relative Policy Optimization (GRPO) have proven highly effective, markedly improving the safety, helpfulness, and utility of generated text. These techniques have recently been extended to speech and audio generation domains. A critical factor in applying RL to audio synthesis is the careful selection of reward functions that accurately reflect perceptual and objective audio quality. In text-to-speech (TTS) systems, RL-based alignment commonly employs rewards based on speaker similarity (SIM) or word error rate (WER). These metrics leverage speech encoders (e.g., WavLM) to capture acoustic properties such as prosody, intonation, pitch, and pauses, while automatic speech recognition models ensure content consistency and intelligibility. For text-to-audio (T2A) generation, suitable reward signals include CLAP similarity (for text-audio semantic alignment), Kullback–Leibler (KL) divergence (for distributional similarity at the sample level), and Fréchet Audio Distance (FAD) (for overall distributional fidelity). In our GRPO-based enhancement pipeline, we adopt these three metrics as primary rewards. GRPO refines the pretrained generation policy by updating it to favor samples that maximize the specified reward. The update relies on an advantage estimate computed from relative rewards within each training batch (or group) of generated samples: the advantage reflects how much better a given output performs compared to the batch average. Following standard practice, we initialize both the policy model and the frozen reference model with the same pretrained weights (e.g., from TangoFlux). Throughout GRPO training, the reference model remains unchanged to provide a stable baseline for advantage computation. We primarily use CLAP similarity, KL divergence, and FAD as reward components, as they collectively capture the most essential dimensions of T2A quality: semantic prompt adherence, sample-level distributional alignment, and global distributional realism.

\paragraph{Reward function}
Our reward function integrates three complementary components—CLAP (for text-audio semantic alignment), FAD (Fréchet Audio Distance, for global distributional fidelity), and KL (Kullback–Leibler divergence, for sample-level distributional similarity)—to comprehensively capture diverse aspects of generated audio quality. We present experimental results demonstrating the training progress and final performance under individual reward configurations as well as their weighted combination.

\begin{equation}
\mathcal{R} = \mathcal{R}_{\text{CLAP}}+\mathcal{R}_{\text{KL}}+\mathcal{R}_{\text{FAD}}
\end{equation}

\paragraph{CLAP}

The CLAP (Contrastive Language–Audio Pretraining) score quantifies the semantic alignment between a generated audio sample and its text prompt by computing the cosine similarity between their embeddings in a shared representation space. This metric effectively captures how well the audio matches the described events, content, and overall intent—analogous to how CLIP (Contrastive Language–Image Pretraining) evaluates text-image alignment in image generation tasks. CLAP is typically pretrained on large-scale audio–text pairs, enabling it to learn a joint embedding space that aligns raw waveforms (or their features) with corresponding natural-language descriptions. Thanks to this contrastive learning objective, CLAP has proven effective across downstream applications, including zero-shot audio classification, automatic audio captioning, and evaluation of text-to-audio (T2A) generation models.

\paragraph{FAD}

The Fréchet Audio Distance (FAD) adapts the Fréchet Inception Distance (FID)—a widely used metric originally developed for image generation—to the audio domain. Like FID, FAD has become a standard objective metric for evaluating the quality of text-to-audio (T2A) and general audio generative models, with a particular focus on measuring perceptual fidelity and distributional similarity to real audio. FAD is computed as follows: embeddings are extracted from both generated and reference (real) audio samples using a pretrained audio classifier (e.g., VGGish, PANNs, or similar). The distributions of these embeddings are then modeled as multivariate Gaussians, and the Fréchet distance (also known as the 2-Wasserstein distance between Gaussians) is calculated between the two distributions. Lower FAD values indicate that the generated audio better matches the reference set in terms of both fidelity and diversity.

\paragraph{KL}

The Kullback–Leibler (KL) divergence serves as a semantic consistency metric in text-to-audio (T2A) evaluation. For each generated–reference audio pair, embeddings are extracted and passed through an AudioSet-pretrained classifier to obtain softmax probability vectors over semantic classes. The KL divergence is then computed from the reference distribution to the generated distribution, quantifying the information loss when approximating the reference semantics with those of the generated audio. Because KL operates at the per-sample level and directly measures mismatch in predicted semantic label distributions, it effectively captures whether the generated audio faithfully preserves the fine-grained content, events, and concepts described by the text prompt (or present in any conditioning signal), complementing global distributional metrics such as FAD.

\section{Experiment}
\label{headings}

\subsection{Training details}

For training, we used the AdamW optimizer in conjunction with a cosine annealing learning rate scheduler to ensure stable and efficient convergence. The initial learning rate was set to 1e-4 for both the data-augmentation pretraining stage and the subsequent GRPO fine-tuning stage, with cosine decay applied to gradually reduce it to a minimum value of 5e-6. All experiments were conducted on NVIDIA A100 GPUs. We utilize an internal audio dataset as the source for data augmentation and reinforcement learning experiments.

\subsection{Evaluation}

\subsubsection{Metrics Analysis}

To evaluate model performance, we employ several objective metrics that assess different aspects of text-to-audio (T2A) generation quality. For each experiment, we report three key scores: Fréchet Distance (FD), Kullback–Leibler divergence (KL), and CLAP score.
\begin{itemize} 
\item Fréchet Distance (FD) measures the distributional similarity between generated and ground-truth audio samples. It is computed using features extracted from a pre-trained audio classifier (e.g., PANNs, VGGish, or others); different backbone classifiers yield different FD values, reflecting variations in feature sensitivity.
\item Kullback–Leibler divergence (KL) quantifies the divergence between the probability distributions of generated and reference audios, typically at the single-sample level. Lower KL values indicate that the generated audio more closely aligns with the ground truth in terms of distributional characteristics.

\item CLAP score computes the cosine similarity between text prompt embeddings and audio embeddings (obtained via a contrastive language-audio pretraining model such as CLAP). Higher scores reflect stronger semantic alignment between the generated audio and the input text prompt.
\end{itemize}

These metrics collectively provide a comprehensive assessment of distributional fidelity (FD), sample-level similarity (KL), and text-audio relevance (CLAP).

\begin{table}[]
\centering
\setlength{\tabcolsep}{20pt}
\begin{tabular}{cccc}
\toprule
Model              & $\text{FD}_{\text{Openl3}}$ ↓    & $\text{KL}_{\text{PaSST}}$ ↓       & CLAP score ↑    \\
\midrule
%ElevenLabs   & 170.78 ± 6.42 & 1.1500 ± 0.0183 & 0.5781 ± 0.0301 \\
TangoFlux  & 181.49 ± 5.66      & 2.1420 ± 0.0202          & 0.3259 ± 0.0254          \\
+dataAug          & 161.93 ± 5.52 & 1.0279 ± 0.0110 & 0.5456 ± 0.0342 \\
 +wtRewd & 158.76 ± 5.74 & 1.0117 ± 0.0192 & 0.5673 ± 0.0299 \\

\bottomrule
\end{tabular}
\caption{Comparing result of baseline TangoFlux model, TangoFlux with data augmentation model, TangoFlux with weighted GRPO (wtRewd) and data augmentation model.}
\label{tab:table2}
\end{table}

Table \ref{tab:table2} presents comparative evaluation results across three model variants: baseline TangoFlux, TangoFlux with data augmentation for GRPO experiment, TangoFlux with weighted GRPO (wGRPO), which combines CLAP, FAD, and KL rewards. Compared to the baseline TangoFlux, the data-augmentation variant consistently outperforms across the FAD, KL, and CLAP metrics. The weighted GRPO (wGRPO) variant yields further modest yet consistent gains over the data-augmentation baseline. Table \ref{tab:table3} reports performance when using single reward functions within GRPO. As anticipated, the CLAP-only reward produces the strongest results on the CLAP metric, since the optimization directly targets this objective, the same for FAD and KL reward. This confirms that single-objective GRPO can reliably enhance the corresponding evaluation metric. While the weighted GRPO configuration achieves performance that is comparable to or superior across all metrics (FAD, KL, CLAP), demonstrating the effectiveness of multi-reward weighting for improving overall text-to-audio generation quality.

\begin{table}[]
\centering
\setlength{\tabcolsep}{20pt}
\begin{tabular}{cccc}
\toprule
Model              & $\text{FD}_{\text{Openl3}}$ ↓    & $\text{KL}_{\text{PaSST}}$ ↓       & CLAP score ↑    \\
\midrule
TangoFlux          & 161.93 ± 5.52 & 1.0279 ± 0.0110 & 0.5456 ± 0.0342 \\
\quad +CLAP   & 171.78 ± 5.45 & 1.0600 ± 0.0144 & 0.5652 ± 0.0329 \\
\quad +FAD   & 157.72 ± 5.31 & 1.0900 ± 0.0176 & 0.5531 ± 0.0271 \\
\quad +KL & 167.45 ± 6.42 & 1.0210 ± 0.0183 & 0.5588 ± 0.0297 \\
\quad +wtRewd & 158.76 ± 5.74 & 1.0117 ± 0.0192 & 0.5673 ± 0.0299 \\
\bottomrule
\end{tabular}
\caption{Comparing result of baseline TangoFlux and its RL version with CLAP, FAD, KL, and wtRewd (weighted rewards) as rewards.}
\label{tab:table3}
\end{table}

\section{Conclusion}

In this study, we investigate Group Relative Policy Optimization (GRPO) as a reinforcement learning approach for text-to-audio (T2A) generation. We systematically explore a range of reward functions—including CLAP, KL divergence, Fréchet Distance (FD), and their weighted combination—and analyze their effects on key evaluation metrics (FAD, KL, CLAP). Our results show that GRPO, especially when using multi-reward weighting, substantially improves overall generation quality by effectively balancing distributional fidelity (FD), sample-level semantic consistency (KL), and prompt adherence (CLAP). Future work will prioritize scaling the training dataset size to enhance model robustness and performance. We intend to conduct controlled experiments examining how increased data volume affects GRPO training dynamics, convergence stability, sample efficiency, and final audio quality in T2A synthesis.

%\section*{References}
%\newpage
%\bibliographystyle{rusnat}
%\bibliography{mybib}

\bibliography{mybib}
\bibliographystyle{plainnat}

%%%%%%%%%%%%%%%%%%%%%%%%%%%%%%%%%%%%%%%%%%%%%%%%%%%%%%%%%%%%

%%% END INSTRUCTIONS %%%

\end{document}